\documentclass[journal]{IEEEtran}

\usepackage{amsmath,amsfonts,amssymb}
\usepackage{array}
\usepackage{balance}
\usepackage{bm}
\usepackage{booktabs}
\usepackage{graphicx}
\usepackage{multirow}
\usepackage{url}
\usepackage{xcolor}
\usepackage{xspace}
\usepackage[subrefformat=parens]{subcaption}
\usepackage[hidelinks]{hyperref}
\usepackage{cleveref}
\usepackage{arydshln}

\newcommand{\bhline}[1]{\noalign{\hrule height #1}}   

\renewcommand{\baselinestretch}{1}
\crefname{equation}{Eq.}{Eqs.}
\crefname{figure}{Fig.}{Figs.}
\crefname{table}{Table}{Table}
\def\BibTeX{{\rm B\kern-.05em{\sc i\kern-.025em b}\kern-.08em T\kern-.1667em\lower.7ex\hbox{E}\kern-.125emX}}

\makeatletter
\let\oldthebibliography\thebibliography
\let\endoldthebibliography\endthebibliography
\renewenvironment{thebibliography}[1]{%
  \oldthebibliography{#1}%
  \setlength{\itemsep}{0.4\baselineskip}
  \setlength{\parsep}{0pt}%
  \setlength{\parskip}{0pt}%
}{\endoldthebibliography}\makeatother

\makeatletter
\let\orig@thebibliography\thebibliography
\let\endorig@thebibliography\endthebibliography
\renewenvironment{thebibliography}[1]{%
  \orig@thebibliography{#1}%
  \fontsize{9pt}{9pt}\selectfont 
}{\endorig@thebibliography}
\makeatother

\newcommand{\method}{AnomSep\xspace}
\newcommand{\projectpage}{\href{https://kokieto.github.io/AnomSep/}{\textit{project page}}}

\begin{document}
\bstctlcite{NoUseDashForRepeatedNames}

\title{\method: Bridging Optimization-Based Separation \\ and Deep Neural Refinement for Anomaly Detection \\ with Listenable Explanations}

\author{
    Koki Shoda$^{\dagger}$,
    Jun Younes Louhi Kasahara$^{\dagger}$,
    Qi An$^{\dagger}$,
    and Atsushi Yamashita$^{\dagger}$ \\
\thanks{
    This work was supported in part by JSPS KAKENHI Grant Numbers JP26KJ0923 and JP26K21348 and by the World-leading Innovative Graduate Study Program for Proactive Environmental Studies (WINGS-PES) at the University of Tokyo.
}
\thanks{Corresponding author: Koki Shoda.}
\thanks{$^{\dag}$ The University of Tokyo 7-3-1 Hongo, Bunkyo-ku, Tokyo 113-8656, Japan.}
\thanks{Digital Object Identifier (DOI): see top of this page.}
}

\maketitle
\markboth{}{}

\begin{abstract}
We propose \method{}, which combines optimiza-tion-based separation with deep neural refinement to achieve accurate anomaly detection with listenable explanations.
Anomaly detection requires no predefined anomalous categories, and understanding the basis for each detection helps security personnel assess the validity of each detection.
End-to-end anomaly detectors can achieve high detection accuracy but output only scalar scores, and existing anomalous sound separation methods provide listenable explanations with limited detection accuracy.
\method{} first obtains preliminary normal-sound and anomalous-sound estimates through optimization-based separation using a dictionary learned from normal sounds in the target environment.
Deep neural refinement then improves the preliminary estimates in a pretrained audio latent space.
The separated anomalous sound serves as the listenable explanation, and its energy determines the anomaly score.
Experiments on synthetic mixtures and real mixtures show improved separation quality over baseline separators.
\method{} achieves detection performance competitive with the end-to-end baseline and provides listenable explanations directly linked to each detection decision.
\end{abstract}

\section{Introduction}
\label{sec:introduction}

\IEEEPARstart{S}{ecurity} monitoring aims to detect events that deviate from everyday conditions and may require the attention of security personnel.
Camera-based monitoring is widely used in security applications~\cite{audio_surveillance_review}.
Acoustic sensing is also useful as a complement to camera-based monitoring because sound can reveal events outside a camera's field of view, behind obstacles, or under poor lighting conditions~\cite{barchiesi2015acoustic,mesaros2018multi}.
A monitoring system that screens long audio streams for anomalous events that differ from normal sounds can therefore provide information unavailable from visual sensing alone.

A key practical challenge is that deviations from everyday conditions are rare and diverse, making it difficult to explicitly define every category of anomaly.
For example, classifying sounds with a zero-shot model~\cite{clap} requires predefined text prompts that serve as class labels for normal and anomalous sounds.
Such closed-set classification may miss anomalies outside the predefined taxonomy.
An alternative is to learn the normal sound distribution in a target environment and detect deviations from that distribution~\cite{novelty_detection_review}.
In this study, an \emph{anomalous sound} is a sound that deviates from the normal sound distribution in the target environment.

\begin{figure}[t]
    \centering
    \includegraphics[width=\columnwidth]{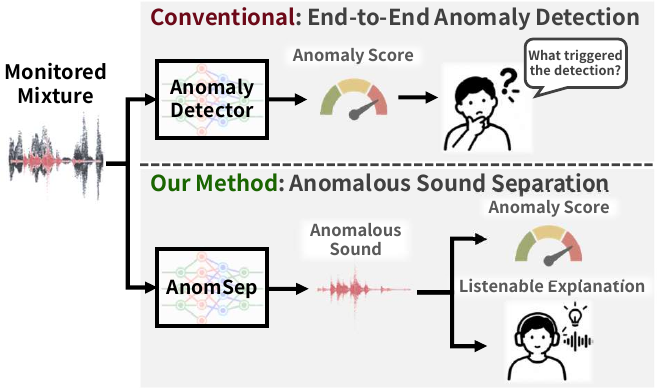}
    \caption{Comparison of end-to-end anomaly detection (top) and anomalous sound separation with \method{} (bottom).
    The end-to-end detector outputs only an anomaly score.
    \method{} presents only the estimated anomalous sound separated from the input mixture as a listenable explanation.
    The mean squared amplitude of the presented waveform determines the anomaly score.}
    \label{fig:motivation}
\end{figure}

Monitored recordings can contain multiple overlapping sounds.
Because false detections are difficult to eliminate completely from monitoring systems, security personnel need to identify which sound triggered each detection to assess the validity of the detection.
An anomaly score alone does not identify the sound responsible for a detection, as shown in \cref{fig:motivation}.
Presenting an anomalous sound separated from the input mixture as a waveform would allow security personnel to hear what was detected.
We adopt the term \emph{listenable explanations} from prior work~\cite{audioLIME} to describe the estimated anomalous-sound waveforms presented to explain individual detections.

End-to-end anomaly detectors~\cite{anomalous_sound_detection_survey, koizumi2020dcase} can achieve high detection accuracy, but their output is a scalar anomaly score and does not provide listenable explanations.
Anomalous sound separation methods~\cite{robust_novelty_NMF,NNE} aim to extract anomalous-sound waveforms that can serve as listenable explanations.
However, existing anomalous sound separation has limited separation performance for normal sounds such as voice because their continuously varying spectra are difficult to represent with a low-rank dictionary~\cite{makishima2019idlma}.
The resulting normal-sound leakage can cause false detections when the energy of the estimated anomalous sound is used as the anomaly score.

The objective of this study is to achieve accurate anomaly detection with listenable explanations.
To achieve this objective, we propose \method{}, which combines optimization-based separation with deep neural refinement.
Figure \ref{fig:motivation} contrasts end-to-end anomaly detection with anomalous sound separation for the same input mixture.
In \method{}, the estimated anomalous-sound waveform provides a listenable explanation, and its energy determines the anomaly score.

The contributions of this study are summarized as follows:
\begin{enumerate}
    \item \textbf{Anomalous sound separation with deep neural refinement:}
    We propose \method{}, which uses optimization-based separation to initialize flow matching in a pretrained audio latent space.
    The neural model refines estimates obtained from a normal-sound dictionary, combining an explicit separation rule with the representational capacity of deep learning.

    \item \textbf{Selection of surrogate anomalous audio clips:}
    We propose Normal Region Exclusion (NRE), which selects surrogate anomalous audio clips from an external audio collection by excluding clips close to normal sounds in an audio embedding space.
    The selection uses the normal sound distribution of the target environment to reduce conflicting supervision during adaptation.

    \item \textbf{Anomaly detection with listenable explanations:}
    We demonstrate that \method{} achieves accurate anomaly detection together with listenable explanations.
    The anomaly score is computed from the same estimated anomalous-sound waveform presented as the listenable explanation, directly linking the waveform to the detection decision.
\end{enumerate}

Audio samples of the separation results are available on the \projectpage\footnote{Project page: \url{https://kokieto.github.io/AnomSep/}}.
The \method{} and NRE implementations are provided in the \href{https://github.com/kokieto/AnomSep}{\textit{repository}}\footnote{Repository: \url{https://github.com/kokieto/AnomSep}}, and the trained \method{} weights are available on Hugging Face\footnote{Model weights: \url{https://huggingface.co/kokieto/AnomSep}}.

\section{Related Work}
\label{sec:related_work}

\subsection{Listenable Explanations for Audio Models}
A review of explanations for audio model predictions surveys methods that visualize relevant features or present listenable explanations~\cite{audio_xai_review}.
For singing voice detection, parts of an audio clip are removed in different combinations to examine how the classifier's prediction changes~\cite{soundlime}.
A linear model is fitted to the resulting predictions to estimate the contributions of temporal segments, frequency components, or time--frequency regions.
Building on prior work on explanations for audio model predictions, we investigate listenable explanations for anomaly detection.

An example of listenable explanations based on source separation is audioLIME~\cite{audioLIME}, which was evaluated on automatic music tagging models.
AudioLIME estimates which separated sounds support the music tagging model's prediction by examining how the prediction changes when sounds are included or excluded.
The separated sounds with larger contributions are presented as listenable explanations.
Because the explanations consist of separated sounds, their quality depends on the separation results; the original paper also identifies limitations on the types of sources that can be separated and separation artifacts as challenges.

In this study, we use the term \emph{environmental sounds} to refer to any sounds monitored for security purposes, including voice.
For environmental sounds, a challenge is that generic source separation does not necessarily yield separate estimates of normal sounds and anomalous sounds in the target environment.

For environmental sound classification, several methods present sounds that support the predicted class as listenable explanations.
A framework based on Nonnegative Matrix Factorization (NMF)~\cite{parekh2024audio_interpretability} reconstructs these sounds by combining relevant NMF components.
Listenable Maps for Audio Classifiers (L-MAC)~\cite{paissan2024listenable} learns to select the time--frequency regions of the input spectrogram that support the predicted class.

In the NMF-based framework~\cite{parekh2024audio_interpretability}, and L-MAC~\cite{paissan2024listenable}, a classifier first produces a prediction, and sounds are then selected or reconstructed to explain the prediction.
The waveform presented as the listenable explanation is an estimate of the audio supporting the prediction, and the classifier is not guaranteed to reproduce the original prediction when given that waveform.
Thus, listening only to the explanations generated by these methods~\cite{parekh2024audio_interpretability, paissan2024listenable} may be insufficient to assess the acoustic basis of the original prediction.

A key feature of \method{} is that the waveform presented as the listenable explanation directly determines the anomaly score.
We first separate anomalous sounds from the input mixture using normal sounds from the target environment as the reference, and then compute the anomaly score as the energy of the estimated anomalous-sound waveform.
Security personnel can therefore listen to the estimated anomalous sound used to compute the score for each detection.

\subsection{End-to-End Anomaly Detection}
Autoencoder-based anomaly detectors are trained to reconstruct normal observations and use the reconstruction error as the anomaly score~\cite{koizumi2020dcase}.
Recent advances in pretrained audio representations~\cite{AudioCLIP, PE_AV} have enabled anomaly detection based on distances between the embeddings of normal audio clips and the embedding of an input audio clip.
By comparing observations with a learned distribution of normal sounds, these methods can provide accurate anomaly detection.
The standard output is a reconstruction error or an embedding distance; these detectors do not directly provide an estimated anomalous-sound waveform from which the anomaly score is computed.
The standard outputs alone do not provide listenable explanations directly linked to the anomaly score.

\subsection{Anomalous Sound Separation}
Nonnegative Novelty Extraction (NNE)~\cite{NNE} has properties complementary to those of end-to-end anomaly detectors.
NNE fixes a nonnegative dictionary learned from normal sounds and decomposes the input magnitude spectrogram into a dictionary reconstruction representing normal sounds and a residual representing anomalous sounds.
NNE can provide listenable explanations because the anomalous sound whose energy determines the anomaly score can be reconstructed as a waveform that security personnel can play back.
However, a fixed low-rank dictionary may not adequately represent complex normal sounds.
Here, \emph{complex normal sounds} refers to sounds such as voice whose spectral variation is difficult to approximate using a small set of fixed spectral bases.
Low-rank NMF models can capture recurring spectral patterns in musical instrument sounds~\cite{yatabe2021hva}.
In contrast, the continuously varying spectra of voice can be difficult to model with low-rank NMF, as discussed in prior work~\cite{makishima2019idlma}.
In NNE, normal-sound content that is poorly represented by the dictionary can remain in the residual, increasing the energy of the anomalous-sound estimate and causing false detections.

\subsection{Generic Source Separation}
Text-prompt-based methods~\cite{audiosep,soloaudio} can separate sounds in a zero-shot setting by specifying which sounds to extract or remove.
CLAPSep~\cite{clapsep} estimates a time--frequency mask for separation through prompt conditioning.
SAM-Audio~\cite{sam_audio} generates separated sounds conditioned on the mixture using flow matching~\cite{FlowMatching}.
Text prompts can describe semantic sound categories such as ``sirens''.
However, such prompts alone may be insufficient to specify the normal sound distribution in the target environment.
Consequently, a generic zero-shot neural separator may have difficulty separating normal sounds and anomalous sounds accurately when applied without adaptation to the target environment.
Nevertheless, pretrained models provide a useful foundation for adaptation to the target environment.

\section{Proposed Method}
\label{sec:method}

\subsection{Concept}
\label{subsec:concept}

\begin{figure*}[t]
    \centering
    \includegraphics[width=\textwidth]{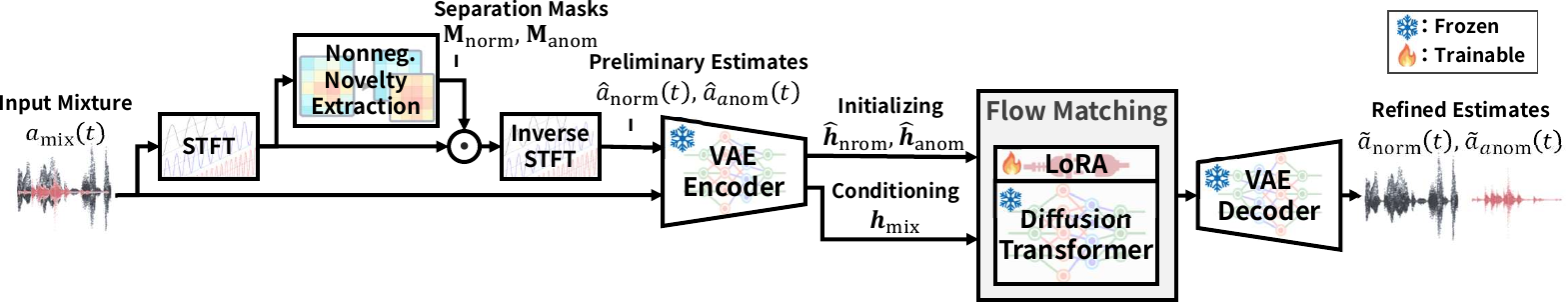}
    \caption{
        Architecture of the proposed \method{}.
        NNE uses a dictionary learned from normal sounds to estimate separation masks and reconstruct preliminary normal-sound and anomalous-sound waveforms.
        The frozen VAE encoder maps the preliminary estimates into latent representations that initialize flow matching, with the mixture latent sequence providing conditioning.
        The Diffusion Transformer with LoRA refines the joint latent sequence, and the frozen VAE decoder reconstructs the refined waveforms.
    }
    \label{fig:over_arch}
\end{figure*}

To provide listenable explanations, we separate anomalous sounds from a mixture and use the energy of the estimated anomalous-sound waveform as the anomaly score.
With this design, the estimated waveform that security personnel listen to is identical to the estimated waveform used to compute the anomaly score.

Optimization-based anomalous sound separation using dictionary learning extracts sounds that cannot be represented by a dictionary learned exclusively from normal sounds, so anomalous-sound categories do not need to be specified in advance.
However, the dictionary has limited representational capacity; if the dictionary cannot adequately represent spectral variation in normal sounds, part of the normal sound is extracted as an anomalous sound, reducing separation accuracy.
For deep-learning-based source separation, a separator with high representational capacity could be adapted to anomalous sound separation using mixtures of normal audio clips and diverse audio clips from an external audio collection.
However, training data cannot easily cover the diversity of anomalous sounds, and learning the separation rule solely from data poses a challenge for generalization to sounds absent from training.

We therefore aim to improve separation performance for diverse anomalous sounds by starting from an explicit separation rule based on a normal-sound dictionary and using the high representational capacity of deep learning to refine the separation results.
We propose \method{}, which uses NNE~\cite{NNE} to obtain preliminary estimates and flow matching in a pretrained audio latent space for deep neural refinement.
SAM-Audio~\cite{sam_audio}, a standard separation method based on flow matching~\cite{FlowMatching}, transforms Gaussian noise into latent representations of separated sounds through a learned velocity field conditioned on the mixture and a prompt.
In contrast, \method{} encodes the normal-sound and anomalous-sound waveforms estimated by NNE into separate latent representations and uses their concatenation as the initial state for flow matching.

Using mixtures of normal audio clips and surrogate anomalous audio clips, the neural network learns a velocity field from the latent representations of the preliminary NNE estimates toward those of the ground-truth normal sounds and anomalous sounds.
This design assigns the neural network the role of correcting normal-sound leakage and missing anomalous-sound content, starting from preliminary estimates obtained using the normal sounds of the target environment.
By using an initial state obtained from an explicit separation rule, we aim to reduce dependence on the training data and promote generalization to diverse anomalous sounds.

\subsection{Architecture}
\label{subsec:architecture}

Figure \ref{fig:over_arch} illustrates the architecture of \method{}.
The system first obtains preliminary estimates using NNE and then produces refined estimates through flow matching in a pretrained audio latent space.
Let $a_{\mathrm{mix}}(t)$ denote the mixture waveform, where $t$ is the discrete-time sample index.
The Short-Time Fourier Transform (STFT) converts the mixture into a complex spectrogram $\mathbf{X}$.
Given a fixed nonnegative dictionary $\mathbf{W}$ learned from normal audio clips, NNE approximates the mixture magnitude spectrogram as
\begin{equation}
    |\mathbf{X}| \approx \mathbf{W}\mathbf{H}+\mathbf{R},
    \label{eq:nne_decomposition}
\end{equation}
where $|\mathbf{X}|$ denotes the element-wise magnitude of $\mathbf{X}$, $\mathbf{H}$ is the nonnegative activation matrix, and $\mathbf{R}$ is the nonnegative residual matrix.
The product $\mathbf{W}\mathbf{H}$ represents the normal-sound magnitude spectrogram reconstructed from the dictionary, and the residual $\mathbf{R}$ is treated as the anomalous-sound magnitude spectrogram.

To reconstruct preliminary estimates for deep neural refinement, we apply Wiener-like soft masks~\cite{fitzgerald2012masking} computed from the NNE magnitude estimates.
We use the squared magnitudes as power estimates and define the normal-sound and anomalous-sound separation masks $\mathbf{M}_{\mathrm{norm}}$ and $\mathbf{M}_{\mathrm{anom}}$, respectively, as
\begin{align}
    \mathbf{M}_{\mathrm{norm}}
    &= \frac{(\mathbf{W}\mathbf{H})^2}
    {(\mathbf{W}\mathbf{H})^2+\mathbf{R}^2},
    \label{eq:normal_mask} \\
    \mathbf{M}_{\mathrm{anom}}
    &= \mathbf{1}-\mathbf{M}_{\mathrm{norm}},
    \label{eq:anom_mask}
\end{align}
where $\mathbf{1}$ denotes a matrix of ones with the same dimensions as $\mathbf{M}_{\mathrm{norm}}$, and squaring, division, and subtraction are applied element-wise.
We apply the masks to the complex mixture spectrogram and use the inverse STFT to reconstruct the preliminary estimates:
\begin{equation}
    \hat{a}_{c}(t)
    = \operatorname{STFT}^{-1}\!\left[\mathbf{M}_{\mathrm{c}} \odot \mathbf{X}\right],
    \label{eq:preliminary_estimates}
\end{equation}
where $c\in\{\mathrm{norm},\mathrm{anom}\}$ identifies the normal sound or anomalous sound, $\odot$ denotes element-wise multiplication, and $\operatorname{STFT}^{-1}$ denotes the inverse STFT.
The hats distinguish the preliminary estimates from the refined estimates, which are denoted by tildes.

For refinement, we use the pretrained Variational Autoencoder (VAE) and Diffusion Transformer (DiT) of SAM-Audio~\cite{sam_audio}.
The frozen VAE encoder $E_{\mathrm{VAE}}$ maps each preliminary estimate and the mixture into latent sequences:
\begin{align}
    \hat{\bm{h}}_c &= E_{\mathrm{VAE}}[\hat{a}_c(t)],
    \label{eq:preliminary_latents} \\
    \bm{h}_{\mathrm{mix}} &= E_{\mathrm{VAE}}[a_{\mathrm{mix}}(t)].
    \label{eq:mixture_latent}
\end{align}

Here, $\hat{\bm{h}}_c$ is the preliminary latent sequence of $\hat{a}_c(t)$, and $\bm{h}_{\mathrm{mix}}$ is the mixture latent sequence used for conditioning.
The initial state for flow matching is
\begin{equation}
    \bm{h}(0)
    = \operatorname{concat}\!\left[\hat{\bm{h}}_{\mathrm{norm}},\hat{\bm{h}}_{\mathrm{anom}}\right],
    \label{eq:nne_initial_state}
\end{equation}
where $\operatorname{concat}$ denotes concatenation along the latent channel dimension.
The joint latent sequence $\bm{h}(\tau)$ evolves with flow time $\tau\in[0,1]$, which is distinct from waveform time $t$.
The DiT predicts the conditional velocity $\bm{v}_{\theta}(\bm{h}(\tau),\tau,\bm{h}_{\mathrm{mix}})$, where $\theta$ denotes the trainable adaptation parameters.

%

Latent diffusion models underlying Stable Diffusion~\cite{stable_diffusion} perform generative modeling in a compact latent space that captures the underlying structure of the data.
Following this principle, we use the pretrained VAE to encode the preliminary normal-sound and anomalous-sound waveforms into lower-dimensional representations and train the DiT to refine the representations through flow matching.
By learning corrections to acoustic content in this pretrained latent space, we aim to improve generalization to anomalous sounds absent from the training data.

The original SAM-Audio~\cite{sam_audio} DiT learns a velocity field from Gaussian noise to the latent representations of separated sources~\cite{FlowMatching}.
In \method{}, the NNE initial state supplies preliminary estimates of the normal sound and anomalous sound, so the velocity field learns to refine an explicit separation result.
The DiT is adapted from general-purpose source separation to the fixed task of separating normal sounds and anomalous sounds in the target environment.
We use Low-Rank Adaptation (LoRA)~\cite{lora} to train low-rank updates to the DiT modules, keeping the pretrained DiT weights and VAE parameters frozen.
Restricting the trainable parameters is intended to stabilize adaptation and limit overfitting to the available training mixtures.

\subsection{Training and Inference}
\label{subsec:training_inference}

\subsubsection{NRE}
Training requires paired normal-sound and anomalous-sound waveforms, but the anomalous sounds encountered during deployment cannot be collected exhaustively.
We therefore select \emph{surrogate anomalous audio clips} from a diverse external audio collection.
An audio clip from the external audio collection may contain sounds that are normal in the target environment; treating such a clip as an anomalous sound would introduce conflicting supervision.
NRE removes diverse audio clips whose embeddings are close to the embeddings of normal audio clips.

Let $\mathcal{D}_{\mathrm{norm}}$ denote the normal audio clips available for model training and $\mathcal{D}_{\mathrm{div}}$ the diverse audio clips used as training candidates.
Normal clips reserved for tuning and validation are excluded from $\mathcal{D}_{\mathrm{norm}}$.
For filtering, we split $\mathcal{D}_{\mathrm{norm}}$ into a reference set $\mathcal{D}_{\mathrm{ref}}$ and a calibration set $\mathcal{D}_{\mathrm{cal}}$ by source recording.
All clips cut from the same recording remain in the same set, so the reference and calibration sets do not share source recordings.

Let $\bm{x}$ denote the waveform vector of a peak-normalized audio clip.
The pretrained PE\_AV~\cite{PE_AV} audio encoder maps $\bm{x}$ to an embedding $E_{PE}[\bm{x}]$, which we normalize to unit Euclidean norm.
Audio embeddings may contain fine variations in timbre and recording conditions that are unnecessary for identifying the neighborhood of normal sounds.
We apply Principal Component Analysis (PCA)~\cite{PCA} to retain the dominant embedding structure in a compact space before computing distances.
Reducing pretrained embeddings before analyzing their group structure has also been studied in topic discovery and document clustering~\cite{embedding_topic_clustering,document_clustering_pipeline}.
Following this approach, we use PCA to define a compact neighborhood geometry and use normal calibration clips to determine how far the excluded region should extend.

The reduced embedding is
\begin{equation}
    \bm{z}(\bm{x}) = \mathbf{P}\{E_{PE}[\bm{x}]-\bm{\mu}\},
    \label{eq:filter_embedding}
\end{equation}
where $\mathbf{P}$ is the PCA projection matrix and $\bm{\mu}$ is the mean embedding for centering.
We fit $\mathbf{P}$ and $\bm{\mu}$ using the normal reference clips and all diverse training candidates, excluding the normal calibration clips.
Whether an external audio clip is suitable as a surrogate anomalous audio clip depends on the normal sounds in the target environment.
We therefore measure proximity to normal sounds using the mean distance to the $k$ nearest normal reference embeddings:
\begin{equation}
    d(\bm{x}) = \frac{1}{k}\sum_{j=1}^{k}
    \left\|\bm{z}(\bm{x})-\bm{z}_j(\bm{x})\right\|_2,
    \label{eq:normal_neighbor_distance}
\end{equation}
where $\bm{z}_j(\bm{x})$ is the $j$th nearest embedding from $\mathcal{D}_{\mathrm{ref}}$, $k$ is the number of neighbors, and $\|\cdot\|_2$ denotes the Euclidean norm.
Distances to the normal reference embeddings vary even among normal clips.
We therefore calibrate the exclusion threshold using an upper quantile of the normal calibration distances and include a margin to reduce the selection of external clips near the normal boundary:
\begin{equation}
    \eta = Q_q + \lambda\left(Q_q-Q_{0.50}\right),
    \label{eq:normal_region_threshold}
\end{equation}
where $\eta$ is the exclusion threshold, $Q_q$ is the empirical $q$-quantile of $\{d(\bm{x}):\bm{x}\in\mathcal{D}_{\mathrm{cal}}\}$, $q\in(0.50,1)$ specifies an upper quantile, $Q_{0.50}$ is the median, and $\lambda$ is a nonnegative margin coefficient.
The margin $\lambda(Q_q-Q_{0.50})$ extends the excluded normal region beyond the upper quantile, with $\lambda$ controlling the exclusion width.
External clips with $d(\bm{x})\leq\eta$ are excluded.
The retained surrogate anomalous audio clips form the set
\begin{equation}
    \mathcal{D}_{\mathrm{sur}}
    = \{\bm{x}\in\mathcal{D}_{\mathrm{div}}:d(\bm{x})>\eta\}.
    \label{eq:surrogate_anomalous_audio}
\end{equation}


\subsubsection{Separator Training}
The normal-sound dictionary provides an explicit separation rule, but the resulting NNE estimates can contain normal-sound leakage and miss anomalous-sound content.
To learn corrections to the preliminary estimates, we train the separator in two stages.
First, NMF learns the normal-sound dictionary $\mathbf{W}$ from the magnitude spectrograms of the normal audio clips in $\mathcal{D}_{\mathrm{norm}}$.
Second, we fix $\mathbf{W}$ and train the LoRA parameters of the DiT to refine the NNE estimates using mixtures of normal audio clips and surrogate anomalous audio clips.
For each training pair, the mixture is
\begin{equation}
    a_{\mathrm{mix}}(t) = a_{\mathrm{norm}}(t)+a_{\mathrm{anom}}(t),
    \label{eq:training_mixture}
\end{equation}
where $a_{\mathrm{norm}}(t)$ and $a_{\mathrm{anom}}(t)$ are the normal-sound and surrogate anomalous-sound waveforms after adjusting their relative levels and applying a common peak-normalization scale.
Normal-sound leakage can produce a nonzero anomalous-sound estimate even when no anomalous sound is present.
We therefore include normal-only training examples with a silent anomalous-sound target, setting $a_{\mathrm{anom}}(t)=0$, to teach the neural model to suppress such leakage.
The original normal-sound and anomalous-sound waveforms provide the supervision, and the NNE estimates provide the initial state in \cref{eq:nne_initial_state}.

Let $\bm{h}_c=E_{\mathrm{VAE}}[a_c(t)]$ denote the ground-truth latent sequence. 
The target joint latent sequence is
\begin{equation}
    \bm{h}_{\mathrm{gt}}
    = \operatorname{concat}\!\left[\bm{h}_{\mathrm{norm}},\bm{h}_{\mathrm{anom}}\right],
    \label{eq:target_latent}
\end{equation}
where the subscript $\mathrm{gt}$ denotes the ground truth.
To train the neural model specifically to correct NNE separation errors, we construct a flow-matching path from the latent representations of the preliminary NNE estimates to those of the ground-truth waveforms.
We sample flow time $\tau$ uniformly from $[0,1]$ and construct the interpolated latent sequence
\begin{equation}
    \bm{h}(\tau)=(1-\tau)\bm{h}(0)+\tau\bm{h}_{\mathrm{gt}}.
    \label{eq:flow_interpolation}
\end{equation}

The corresponding target velocity, $\bm{h}_{\mathrm{gt}}-\bm{h}(0)$, represents the latent correction required for each training mixture.
The LoRA parameters are optimized by minimizing the flow-matching loss
\begin{equation}
    \mathcal{L}_{\mathrm{FM}}
    = \mathbb{E}\!\left[
    \left\|\bm{v}_{\theta}(\tau)
    -\left(\bm{h}_{\mathrm{gt}}-\bm{h}(0)\right)\right\|^2
    \right],
    \label{eq:flow_matching_loss}
\end{equation}
where $\bm{v}_{\theta}(\tau)$ abbreviates $\bm{v}_{\theta}(\bm{h}(\tau),\tau,\bm{h}_{\mathrm{mix}})$, and $\mathcal{L}_{\mathrm{FM}}$ denotes the training objective.
The expectation is over the sampled training mixtures and flow times, and $\|\cdot\|^2$ denotes the mean squared value.
No gradient is propagated through NNE or the VAE.

\subsubsection{Inference and Anomaly Scoring}
At inference, NNE produces the preliminary estimates from the observed mixture, and the VAE encoder constructs $\bm{h}(0)$ and $\bm{h}_{\mathrm{mix}}$.
The joint latent sequence is refined by numerically integrating
\begin{equation}
    \frac{\mathrm{d}\bm{h}(\tau)}{\mathrm{d}\tau}
    = \bm{v}_{\theta}(\bm{h}(\tau),\tau,\bm{h}_{\mathrm{mix}}),
    \label{eq:flow_inference}
\end{equation}
from $\tau=0$ to $\tau=1$, keeping the mixture condition fixed.
The final state is split along the channel dimension into refined normal-sound and anomalous-sound latent sequences $\bm{h}_{\mathrm{norm}}(1)$ and $\bm{h}_{\mathrm{anom}}(1)$.
The frozen VAE decoder $D_{\mathrm{VAE}}$ reconstructs the refined waveforms as $\tilde{a}_c(t)=D_{\mathrm{VAE}}[\bm{h}_c(1)]$ for $c\in\{\mathrm{norm},\mathrm{anom}\}$.

To link each detection decision directly to the listenable explanation, we compute the anomaly score from the same refined anomalous-sound waveform.
Specifically, we use the mean squared amplitude of the waveform:
\begin{equation}
    s
    = \frac{1}{T}\sum_{t=1}^{T}\tilde{a}_{\mathrm{anom}}(t)^2,
    \label{eq:anomaly_score}
\end{equation}
where $s$ is the scalar anomaly score and $T$ is the number of waveform samples in the clip.
A clip is detected as containing an anomalous sound when the score is at least a detection threshold.

\section{Experimental Setup}
\label{sec:exp_setup}

\subsection{Experimental Design Overview}

The experiments evaluated anomalous sound separation and anomaly detection using \emph{synthetic mixtures} and \emph{real mixtures}.
Detection metrics evaluated anomaly detection accuracy, and separation metrics evaluated the quality of the anomalous-sound waveforms presented as listenable explanations.
The direct correspondence between each presented waveform and its anomaly score follows from the scoring rule in \cref{eq:anomaly_score}.
Synthetic mixtures were constructed by adding separately recorded normal sounds and anomalous sounds, providing isolated source references for evaluating the separated anomalous-sound waveforms.
Real mixtures were urban recordings of naturally overlapping sounds obtained without artificial mixing, providing a complementary evaluation under actual recording conditions.
Both evaluation conditions included normal-only clips to assess detection in the absence of anomalous sounds.
We trained a separate model for each target environment because normal sound distributions differ across environments.
We kept the learned parameters and detection thresholds fixed during final evaluation.
All methods were evaluated on the same audio clips within each condition.

\subsection{Datasets and Mixture Construction}

\subsubsection{Synthetic Mixtures}
We used the airport, metro station, and public square scenes from the TAU Urban Acoustic Scenes 2019 development dataset~\cite{TAU2019Development} as three target environments.
Only recordings from device A were used.
For each scene, we assigned 80\% of the source recordings to training and 20\% to final evaluation, and divided each 10~s recording into two nonoverlapping 5~s clips.
We held out 20\% of the training clips for tuning and validation.
The remaining clips were used to learn the normal-sound dictionary and train the models.

The anomalous sounds used for evaluation comprised sirens from ESC-50~\cite{ESC50}, glass breaking, gunshots, and baby cries from TUT Rare Sound Events 2017~\cite{TUTRareSoundEvents2017}, and screams from the Human Screaming Detection Dataset~\cite{HumanScreamingDetectionDataset}.
Anomalous-sound categories were sampled with equal probability.
The evaluation set contained equal numbers of normal-only clips and mixtures of normal sounds and anomalous sounds.
For mixtures containing an anomalous sound, the anomalous-to-normal energy ratio was sampled uniformly from $-10$ to $5$~dB.

We also varied surrogate audio selection for the synthetic-mixture experiments to examine the effects of the NRE region size and robustness to anomalous-sound categories excluded from task-specific training and tuning.
To examine the NRE region size, we varied the margin coefficient over $\lambda\in\{0.0,0.5,1.0,1.5,2.0,2.5,3.0\}$ without category exclusion.
For each margin, the FSD50K clips retained by NRE were used for separator training and validation, NNE dictionary selection, and detection threshold tuning.
The No Processing condition applied no surrogate audio selection, retained the main experiment's dictionary and tuning set, and used unfiltered FSD50K clips for separator training and validation.

The $\lambda=1.0$ condition reused the main experiment's results.
The normal audio splits, training seed, training sample budget, and final evaluation mixtures were fixed across conditions.
For the independently trained NRE conditions, the candidate NNE dictionaries were reused, and the number of atoms and detection threshold were selected using the corresponding tuning mixtures.
The margin was not selected using the final evaluation data.

To examine robustness to the exclusion of evaluation-related categories from task-specific training and tuning, we trained and evaluated a variant of \method{} with the corresponding FSD50K~\cite{FSD50K} categories excluded.
We denote this variant as NRE w/ Category Excl. ($\lambda=1.0$).
We removed every FSD50K clip labeled Siren, Glass, Shatter, Gunshot\_and\_gunfire, Crying\_and\_sobbing, or Screaming from the development and evaluation splits before fitting PCA and applying NRE with $\lambda=1.0$.
The excluded clips were unavailable for separator training and validation, NNE dictionary selection, and detection threshold tuning.
The exclusion rate was the fraction of clips removed from the FSD50K development split by category exclusion and NRE, as applicable.
The rate measures the reduction of the audio clip pool before sampling training examples. This exclusion does not control category exposure during model pretraining.

\subsubsection{Real Mixtures}
We used SONYC-UST~\cite{SONYCUSTV2}, retaining the full 10~s duration of each recording.
Alert signals, including car horns, sirens, car alarms, and reverse beepers, were treated as anomalous sounds.
Using the three citizen annotations for each recording, we defined normal recordings by unanimous absence of alert signals and recordings containing anomalous sounds by unanimous presence.
Recordings with annotation disagreement were excluded.
We constructed recording-disjoint splits for this experiment: all recordings containing anomalous sounds and an equal number of normal recordings were reserved for final evaluation, and the remaining normal recordings were divided into 80\% for model training and 20\% for tuning and validation.
The final evaluation used the observed recordings without artificial mixing.
No recording labeled as containing alert signals was used for model training or tuning.

\subsubsection{Surrogate Anomalous Audio Clips}
For both evaluation conditions, we used FSD50K~\cite{FSD50K} as the external audio collection, drawing audio clips for training from its development split and audio clips for tuning and validation from its evaluation split.
NRE selected surrogate anomalous audio clips separately for each target environment.
Diverse audio clips were cropped or repeated to match the duration of the normal audio clips.
Training and validation examples consisted of a normal audio clip alone with a probability of 20\%; the remaining examples were mixtures of normal audio clips and surrogate anomalous audio clips with an anomalous-to-normal energy ratio sampled uniformly from $-15$ to $5$~dB.
The tuning sets used to select hyperparameters and detection thresholds contained equal numbers of normal-only examples and mixtures of normal audio clips and surrogate anomalous audio clips.

\subsection{Compared Methods}
The detection-only comparisons were the autoencoder-based DCASE2020--2026 Baseline~\cite{koizumi2020dcase} and PE\_AV~\cite{PE_AV} w/ $k$-NN Distance, where $k$-NN denotes $k$-Nearest Neighbor.
The autoencoder was trained on log-mel spectrograms and used reconstruction error as the anomaly score.
PE\_AV w/ $k$-NN Distance used the mean Euclidean distance to the $k$ nearest normal training embeddings as the anomaly score, with all embeddings normalized to unit norm.
For each target environment, we selected $k$ from $\{8,16,32,64,128,256\}$ by maximizing accuracy on the tuning set after optimizing the detection threshold for each candidate.
Ties were resolved in favor of the smaller $k$, and the selected $k$ and threshold were fixed for final evaluation.

The separation-based comparisons were NNE~\cite{NNE} and SAM-Audio~\cite{sam_audio}.
NNE used a dictionary learned from normal audio clips in the target environment to obtain the preliminary estimates without deep neural refinement.
Without task-specific fine-tuning, SAM-Audio extracted the normal sound using a fixed normal-sound prompt, and the residual output was treated as the anomalous sound.
For all separation-based methods, the anomaly score was the mean squared amplitude of the estimated anomalous-sound waveform.
No Processing used the input mixture as the anomalous-sound waveform to provide a reference for separation quality.

We included two comparisons to examine the proposed design.
SAM-Audio w/ Fine-Tuning used the same normal audio clips, surrogate anomalous audio clips, and LoRA training procedure as \method{}.
The initial state was Gaussian noise, allowing us to examine the contribution of NNE initialization.
\method{} w/o NRE used the diverse audio clips directly for separator training and validation.

\subsection{Evaluation Metrics}
\label{subsec:eval_metrics}
We evaluated separation using Contrastive Language--Audio Pretraining (CLAP)~\cite{laion_clap} and SAM Audio Judge (SAJ)~\cite{SAJScore}, following SAM-Audio~\cite{sam_audio}.
For synthetic mixtures, CLAP$_{\mathrm{A}}$ was the cosine similarity between the estimated anomalous-sound waveform and its isolated reference in the CLAP audio embedding space.
For real mixtures, isolated source references were unavailable, so CLAP$_{\mathrm{T}}$ measured text--audio similarity to the description ``Alert signal such as a car horn, siren, car alarm, or reverse beeper.''
We also reported the automated Overall score from SAJ for both evaluation conditions.
Separation scores were averaged over evaluation clips containing anomalous sounds.

We evaluated detection using Accuracy and the F1 score over all evaluation clips.
For each method and target environment, the detection threshold was selected to maximize accuracy on the tuning set and was then fixed for final evaluation.
Thus, detection thresholds were determined using surrogate anomalous audio clips without using the final evaluation labels.
In all result tables, the highest displayed value in each performance column is shown in bold, including ties.

\subsection{Implementation Details}
Audio was processed at 16~kHz and peak-normalized, with resampling to 48~kHz for SAM-Audio.
For NRE, we used the small PE\_AV model, a PCA dimension of 100, $k=32$, $q=0.95$, and $\lambda=1.0$.
With $\lambda=1.0$, the margin added to the upper quantile equals the difference between the upper quantile and the median of the normal calibration distances in \cref{eq:normal_region_threshold}.
The normal clips available for model training were split by source recording into 80\% for reference and 20\% for calibration, and empirical quantiles were computed by linear interpolation.
We used the same PCA projection, normal reference set, and exclusion threshold to select surrogate anomalous audio clips from the FSD50K evaluation split for tuning and validation.

NNE used a Hann window of 64~ms and a hop of 16~ms.
We selected the number of dictionary atoms from $\{2,4,8,16,32,64\}$ by maximizing the mean CLAP similarity on tuning mixtures of normal audio clips and surrogate anomalous audio clips.
The selection used CLAP$_{\mathrm{A}}$ for the experiments with synthetic mixtures and CLAP$_{\mathrm{T}}$ with descriptions derived from the FSD50K labels for the experiment with real mixtures.
The selected dictionary was shared by NNE, \method{}, and \method{} w/o NRE within each target environment.

We trained LoRA with rank 16 using AdamW, an initial learning rate of $1.0\times10^{-3}$, and cosine learning-rate decay.
Batch sizes were 48 for the experiments with synthetic mixtures and 24 for the experiment with real mixtures.
We selected the checkpoint with the lowest flow-matching loss on validation mixtures of held-out normal audio clips and surrogate anomalous audio clips.
For \method{} w/o NRE, the validation mixtures used unfiltered audio clips from the FSD50K evaluation split.
For flow-matching inference in \method{}, we used the midpoint solver with 16 integration steps.
The normal-sound prompt was ``Background noise''.
Inference time was measured on an NVIDIA GeForce RTX 5090 with a batch size of one, using the same 100 evaluation clips across the three synthetic-mixture environments for all methods.

\section{Results and Discussion}
\label{sec:results}

\subsection{Anomalous Sound Separation and Anomaly Detection on Synthetic Mixtures}

\begin{figure*}[!tp] 
    \centering
    \begin{minipage}[t]{0.95\linewidth}
        \centering
        \includegraphics[width=0.91\linewidth]{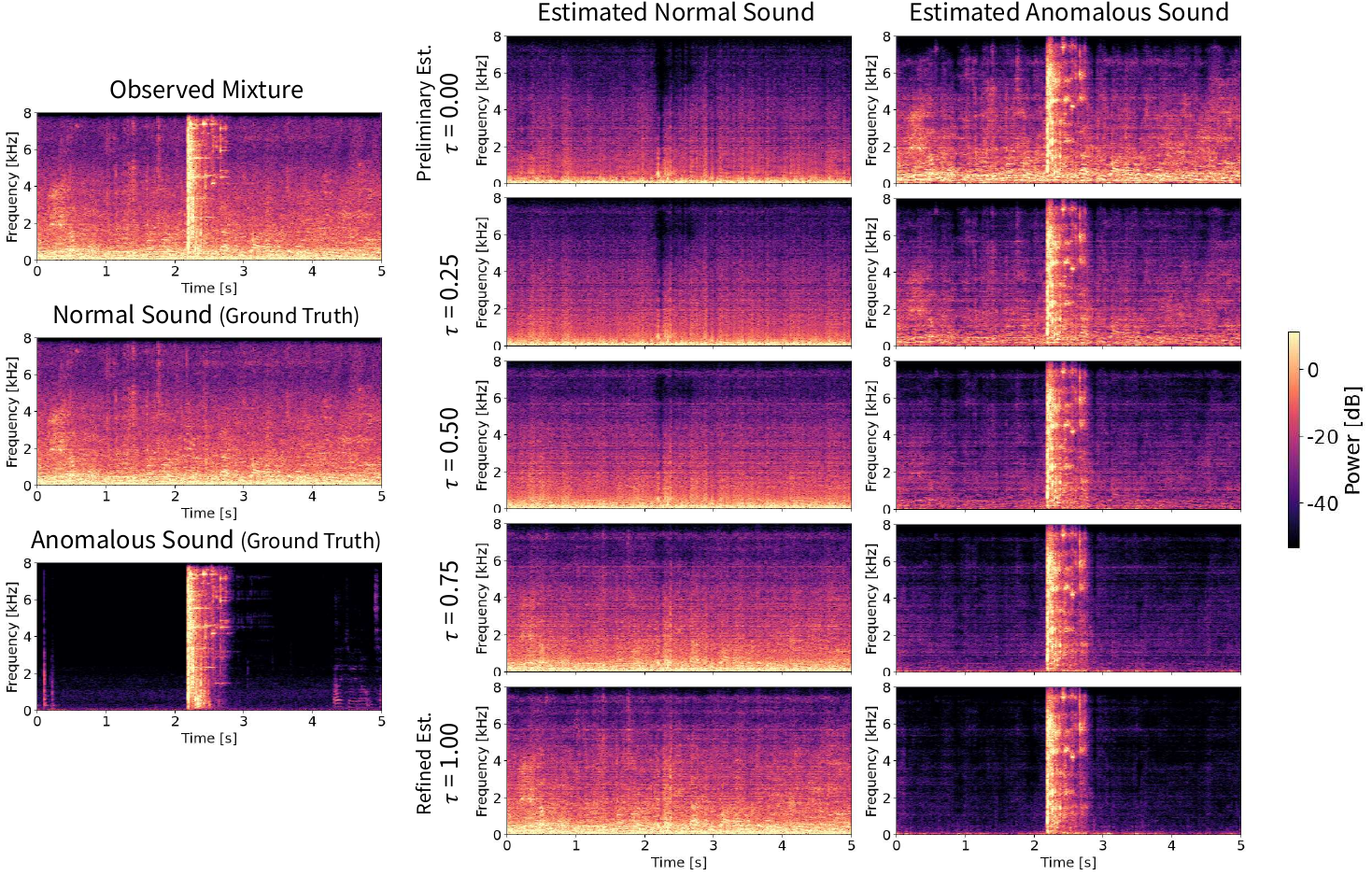}
        \subcaption{
            \footnotesize
            In this example, the normal sound is airport background noise, and the anomalous sound is glass breaking.
            The preliminary anomalous-sound estimate captures the glass-breaking sound but also contains substantial normal-sound leakage.
            As flow time advances, normal-sound leakage decreases, leaving only the glass-breaking sound in the anomalous-sound estimate.
            The preliminary normal-sound estimate retains only part of the normal sound.
            As flow time advances, the airport background noise is progressively restored in the normal-sound estimate.
        }
        \label{fig:sample1}
    \end{minipage}
    \par\vspace{10pt}
    \centering
    \begin{minipage}[t]{0.95\linewidth}
        \centering
        \includegraphics[width=0.91\linewidth]{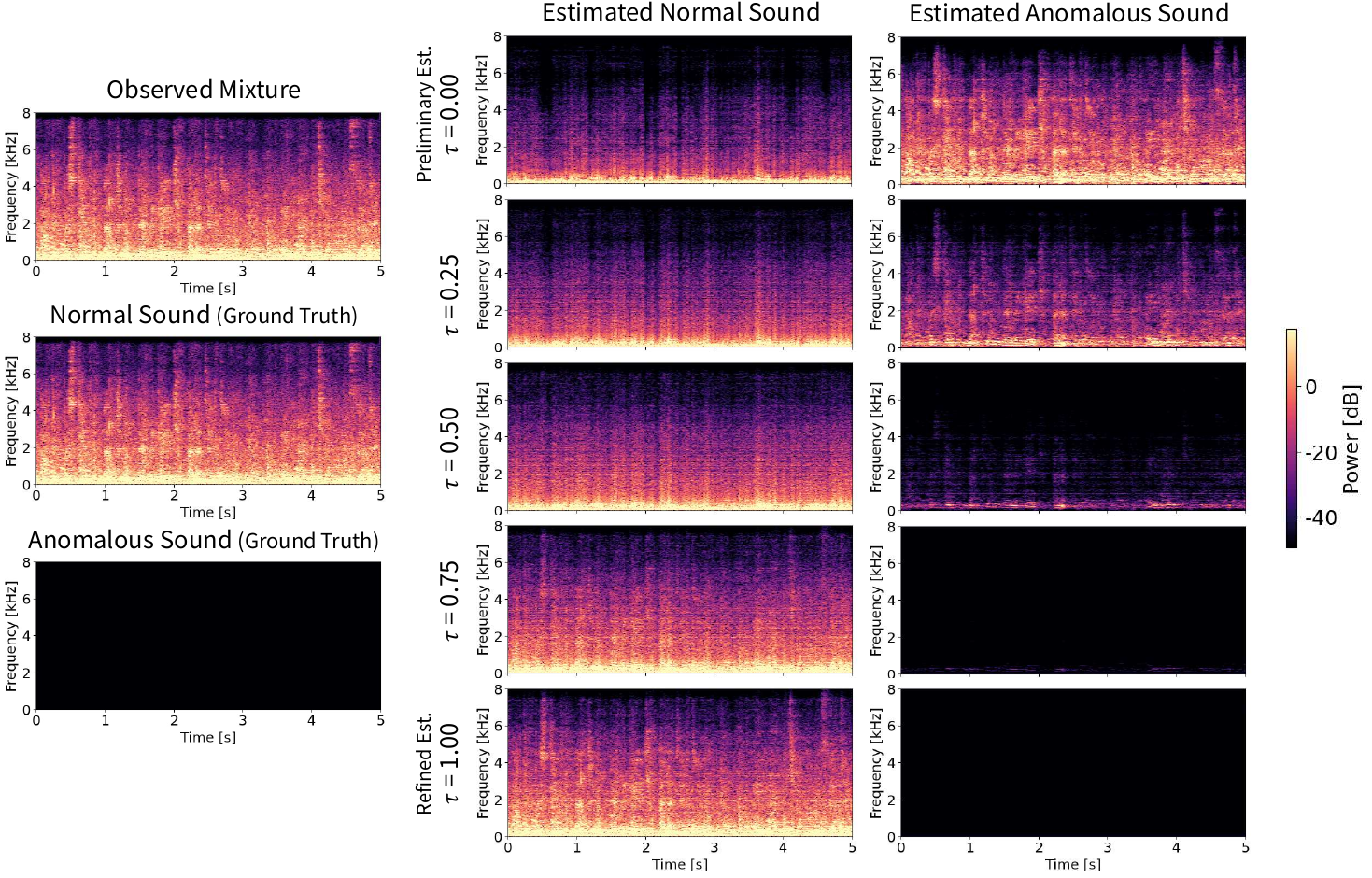}
        \subcaption{
            \footnotesize
            In this example, the normal sound is an airport announcement, and no anomalous sound is present (silence).
            The preliminary anomalous-sound estimate incorrectly captures the announcement.
            As flow time advances, the announcement fades from the anomalous-sound estimate, and the estimate approaches silence.
            The preliminary normal-sound estimate fails to capture the announcement.
            As flow time advances, the announcement is progressively restored in the normal-sound estimate.
        }
        \label{fig:sample2}
    \end{minipage}
    \caption{
        Deep neural refinement process.
        The estimates evolve from the preliminary estimates obtained by NNE (top) to the final refined estimates (bottom) as flow time $\tau$ advances from 0 to 1.
    }
    \label{fig:robot}
\end{figure*}

Figure \ref{fig:robot} shows that deep neural refinement suppresses normal-sound leakage in the NNE estimates and retains anomalous-sound content when present.
The correction in both mixtures and normal-only clips supports the effectiveness of deep neural refinement in obtaining a reliable anomalous-sound waveform for anomaly detection.
Comparisons of separation results with other methods and playable audio samples are available on the \projectpage.

\begin{table*}[t]
    \centering
    \caption{
        Anomalous sound separation and anomaly detection performance of \method{} under different surrogate audio selection settings on synthetic mixtures.
        No Processing applies no surrogate audio selection.
        NRE w/ Category Excl. removes FSD50K clips with labels related to the anomalous-sound categories used for evaluation.
        Excl. [\%] is the percentage of FSD50K development clips removed by NRE and, where applicable, category exclusion.
    }
    \label{tab:surrogate_selection}
    \footnotesize
    \setlength{\tabcolsep}{2pt}
    \begin{tabular}{l|ccccc|ccccc|ccccc}
        \bhline{1pt}
        & \multicolumn{5}{c|}{Airport} & \multicolumn{5}{c|}{Metro station} & \multicolumn{5}{c}{Public square} \\
        Surrogate audio selection
        & Excl. [\%] & CLAP$_{\mathrm{A}}\!\uparrow$ & SAJ$\uparrow$ & Acc.$\uparrow$ & F1$\uparrow$
        & Excl. [\%] & CLAP$_{\mathrm{A}}\!\uparrow$ & SAJ$\uparrow$ & Acc.$\uparrow$ & F1$\uparrow$
        & Excl. [\%] & CLAP$_{\mathrm{A}}\!\uparrow$ & SAJ$\uparrow$ & Acc.$\uparrow$ & F1$\uparrow$ \\
        \hline
        No Processing
        & 0.00 & 0.68 & 3.85 & 0.91 & 0.91
        & 0.00 & 0.71 & 3.75 & 0.86 & 0.87
        & 0.00 & \textbf{0.73} & 4.00 & 0.91 & 0.92 \\
        NRE ($\lambda=0.0$)
        & 0.83 & 0.69 & 3.99 & 0.92 & 0.92
        & 3.50 & \textbf{0.72} & 3.80 & 0.86 & 0.88
        & 1.34 & \textbf{0.73} & 3.94 & 0.89 & 0.90 \\
        NRE ($\lambda=0.5$)
        & 2.58 & \textbf{0.70} & 3.97 & 0.91 & 0.92
        & 8.69 & 0.70 & 3.89 & 0.91 & 0.92
        & 5.13 & \textbf{0.73} & 3.89 & 0.86 & 0.88 \\
        NRE ($\lambda=1.0$)
        & 7.60 & \textbf{0.70} & 3.90 & 0.89 & 0.90
        & 24.99 & 0.71 & \textbf{3.93} & 0.91 & 0.92
        & 18.63 & \textbf{0.73} & 4.03 & 0.92 & 0.93 \\
        NRE ($\lambda=1.5$)
        & 28.44 & 0.69 & 3.93 & 0.92 & 0.93
        & 69.30 & 0.71 & 3.92 & \textbf{0.92} & \textbf{0.93}
        & 74.39 & \textbf{0.73} & 3.97 & \textbf{0.94} & \textbf{0.94} \\
        NRE ($\lambda=2.0$)
        & 80.98 & 0.68 & \textbf{4.00} & \textbf{0.95} & \textbf{0.95}
        & 94.46 & 0.69 & 3.72 & 0.90 & 0.90
        & 97.79 & 0.71 & 4.03 & \textbf{0.94} & \textbf{0.94} \\
        NRE ($\lambda=2.5$)
        & 97.36 & 0.61 & 3.54 & 0.89 & 0.88
        & 99.30 & 0.64 & 3.58 & 0.91 & 0.91
        & 99.85 & 0.66 & 3.75 & 0.88 & 0.87 \\
        NRE ($\lambda=3.0$)
        & 99.76 & 0.60 & 3.57 & 0.89 & 0.89
        & 99.94 & 0.54 & 3.20 & 0.82 & 0.78
        & 100.00 & -- & -- & -- & -- \\
        \hdashline
        NRE w/ Category Excl. ($\lambda=1.0$)
        & 10.43 & 0.69 & 3.96 & 0.91 & 0.91
        & 27.05 & 0.70 & 3.87 & 0.89 & 0.90
        & 21.59 & \textbf{0.73} & \textbf{4.04} & \textbf{0.94} & \textbf{0.94} \\
        \bhline{1pt}
    \end{tabular}
\end{table*}

\Cref{tab:surrogate_selection} shows that moderate NRE exclusion improved detection performance over No Processing in every scene.
More extensive exclusion reduced detection and separation performance.
Removing near-normal clips may reduce ambiguity between normal sounds and surrogate anomalous targets.
Excessive exclusion narrows the diversity of surrogate sounds, so a fixed training budget increasingly repeats the remaining clips and may encourage overfitting to a limited set of acoustic patterns.
The reduced diversity also affects NNE dictionary and detection threshold selection, potentially weakening generalization to evaluation anomalies.

NRE w/ Category Excl. showed no substantial degradation in separation or detection relative to the default NRE setting.
These results suggest that \method{} may generalize to anomalous-sound categories excluded from task-specific training and tuning.

\begin{table*}[t]
    \centering
    \caption{
        Anomalous sound separation and anomaly detection performance on synthetic mixtures.
        No Processing uses the input mixture as the anomalous-sound waveform.
    }
    \label{tab:tau_results}
    \footnotesize
    \setlength{\tabcolsep}{3pt}
    \begin{tabular}{l|cccc|cccc|cccc}
        \bhline{1pt}
        & \multicolumn{4}{c|}{Airport} & \multicolumn{4}{c|}{Metro station} & \multicolumn{4}{c}{Public square} \\
        Method
        & CLAP$_{\mathrm{A}}\!\uparrow$ & SAJ$\uparrow$ & Acc.$\uparrow$ & F1$\uparrow$
        & CLAP$_{\mathrm{A}}\!\uparrow$ & SAJ$\uparrow$ & Acc.$\uparrow$ & F1$\uparrow$
        & CLAP$_{\mathrm{A}}\!\uparrow$ & SAJ$\uparrow$ & Acc.$\uparrow$ & F1$\uparrow$ \\
        \hline
        No Processing
        & 0.55 & 1.38 & - & -
        & 0.62 & 1.49 & - & -
        & 0.61 & 1.50 & - & - \\
        DCASE2020--2026 Baseline~\cite{koizumi2020dcase}
        & - & - & 0.85 & 0.84
        & - & - & 0.86 & 0.87
        & - & - & 0.86 & 0.87 \\
        PE\_AV~\cite{PE_AV} w/ $k$-NN Distance
        & - & - & \textbf{0.90} & \textbf{0.90}
        & - & - & 0.90 & 0.90
        & - & - & \textbf{0.92} & 0.92 \\
        NNE~\cite{NNE}
        & 0.57 & 2.19 & 0.65 & 0.55
        & 0.63 & 2.34 & 0.63 & 0.53
        & 0.63 & 2.60 & 0.67 & 0.64 \\
        SAM-Audio~\cite{sam_audio}
        & 0.39 & 1.83 & 0.50 & 0.67
        & 0.43 & 1.84 & 0.50 & 0.67
        & 0.44 & 1.87 & 0.50 & 0.67 \\
        SAM-Audio~\cite{sam_audio} w/ Fine-Tuning
        & 0.66 & 3.86 & 0.89 & \textbf{0.90}
        & 0.68 & 3.83 & 0.84 & 0.85
        & 0.70 & 3.81 & 0.90 & 0.90 \\
        \method{} (Ours)
        & \textbf{0.70} & \textbf{3.90} & 0.89 & \textbf{0.90}
        & \textbf{0.71} & \textbf{3.93} & \textbf{0.91} & \textbf{0.92}
        & \textbf{0.73} & \textbf{4.03} & \textbf{0.92} & \textbf{0.93} \\
        \bhline{1pt}
    \end{tabular}
\end{table*}

\Cref{tab:tau_results} compares \method{} using the default NRE setting with the other methods on synthetic mixtures.
\method{} achieved the highest displayed detection accuracy and F1 score in the metro station and public square, with accuracy tied with PE\_AV w/ $k$-NN Distance in the public square.
In the airport, \method{} achieved the highest displayed F1 score, tied with PE\_AV w/ $k$-NN Distance and SAM-Audio w/ Fine-Tuning.
These results show that \method{} provides listenable explanations with detection performance competitive with the end-to-end baseline, using the energy of the presented waveform as the anomaly score.

For separation, NNE improved CLAP$_{\mathrm{A}}$ and SAJ over No Processing in every environment, but NNE's detection performance remained below that of the detection-only methods.
SAM-Audio without task-specific fine-tuning produced lower CLAP$_{\mathrm{A}}$ scores than No Processing and lower SAJ scores than NNE across all environments.
The results indicate that a normal-sound dictionary alone or a general-purpose separator with a fixed prompt is insufficient for accurate anomalous sound separation and anomaly detection under the evaluated conditions.
\method{} exceeded NNE and SAM-Audio in both separation metrics and both detection metrics across all environments.

Fine-tuning SAM-Audio improved every separation and detection metric relative to SAM-Audio without task-specific fine-tuning on synthetic mixtures.
\method{} improved both separation metrics over SAM-Audio w/ Fine-Tuning in every environment and improved both detection metrics in the metro station and public square.
The adapted models used the same training data and LoRA training procedure, so the additional improvement supports initializing refinement with the NNE estimates.
Together with the improvement over NNE alone, the comparison supports the complementary roles of optimization-based separation and deep neural refinement.

\subsection{Anomalous Sound Separation and Anomaly Detection on Real Mixtures}
\begin{table}[t]
    \centering
    \caption{
        Anomalous sound separation and anomaly detection performance on real mixtures.
        Recordings containing alert signals were reserved for final evaluation.
        No Processing uses the input mixture as the anomalous-sound waveform.
    }
    \label{tab:sonyc_results}
    \footnotesize
    \setlength{\tabcolsep}{2pt}
    \begin{tabular}{l|cccc}
        \bhline{1pt}
        Method & CLAP$_{\mathrm{T}}\!\uparrow$ & SAJ$\uparrow$ & Acc.$\uparrow$ & F1$\uparrow$ \\
        \hline
        No Processing
        & 0.21 & 1.30 & - & - \\
        DCASE2020--2026 Baseline~\cite{koizumi2020dcase}
        & - & - & 0.67 & 0.59 \\
        PE\_AV~\cite{PE_AV} w/ $k$-NN Distance
        & - & - & 0.57 & 0.30 \\
        NNE~\cite{NNE}
        & 0.20 & 2.13 & 0.58 & 0.45 \\
        SAM-Audio~\cite{sam_audio}
        & 0.15 & 1.74 & 0.50 & \textbf{0.66} \\
        SAM-Audio w/ Fine-Tuning
        & 0.31 & 2.26 & 0.66 & 0.58 \\
        \shortstack[l]{\method{} w/o NRE}
        & 0.32 & 2.40 & 0.66 & 0.58 \\
        \method{} (Ours)
        & \textbf{0.33} & \textbf{2.74} & \textbf{0.69} & \textbf{0.66} \\
        \bhline{1pt}
    \end{tabular}
\end{table}

\Cref{tab:sonyc_results} summarizes the results on real mixtures.
\method{} achieved the highest SAJ and detection accuracy among the compared methods, and its displayed F1 score was tied with SAM-Audio.
Training and threshold selection used normal recordings and surrogate anomalous audio clips, with real mixtures containing alert signals reserved for final evaluation.
The separation and detection gains therefore support transfer from training mixtures constructed using surrogate anomalous audio clips to real mixtures.

All separation-based methods increased SAJ relative to No Processing.
\method{} improved both CLAP$_{\mathrm{T}}$ and SAJ over No Processing.
The accompanying improvement in detection supports using the refined anomalous-sound waveform to compute the anomaly score; the same waveform is available for listening as the listenable explanation.

SAM-Audio w/ Fine-tuning improved both separation metrics and detection accuracy, although the F1 score decreased relative to SAM-Audio w/ Fine-Tuning.
\method{} exceeded SAM-Audio w/ Fine-Tuning in SAJ and both detection metrics.
Relative to \method{} w/o NRE, NRE improved SAJ and both detection metrics.
The SAJ and detection results on real mixtures support selecting surrogate anomalous audio clips with reference to the normal sound distribution of the target environment.

\subsection{Limitations}
\label{sec:limitations}
\begin{table}[t]
    \centering
    \caption{
        Inference time per five-second audio clip (mean $\pm$ standard deviation).
    }
    \label{tab:inference_time}
    \footnotesize
    \begin{tabular}{l|c}
        \bhline{1pt}
        Method & Inference Time [s] \\
        \hline
        DCASE2020--2026 Baseline~\cite{koizumi2020dcase} & $\textbf{0.002} \pm 0.000$ \\
        PE\_AV~\cite{PE_AV} w/ $k$-NN Distance & $0.023 \pm 0.001$ \\
        NNE~\cite{NNE} & $0.025 \pm 0.001$ \\
        SAM-Audio~\cite{sam_audio} & $0.449 \pm 0.004$ \\
        SAM-Audio~\cite{sam_audio} w/ Fine-Tuning & $0.468 \pm 0.004$ \\
        \method{} (Ours) & $0.524 \pm 0.004$ \\
        \bhline{1pt}
    \end{tabular}
\end{table}

As shown in \cref{tab:inference_time}, \method{} processed a five-second clip in $0.524$~s, corresponding to a real-time factor of approximately 0.10.
The additional computation relative to SAM-Audio w/ Fine-Tuning mainly consists of NNE initialization and VAE encoding of the preliminary normal-sound and anomalous-sound estimates to construct the initial latent state.
The faster-than-real-time processing is promising for anomaly detection with listenable explanations under the evaluated GPU setting.
However, \method{} has a higher computational cost than the compared methods, and the timing evaluates complete clips without assessing streaming latency.
A promising direction for reducing the computational cost of neural refinement is to adapt one-step generative approaches, such as MeanFlow~\cite{meanflow}, to the conditional refinement stage.

Another limitation is that the models and detection thresholds were fixed, leaving adaptation to temporal changes in the normal sound distribution unexamined.
Changes in background activity or recording conditions may increase false detections when sounds considered normal by security personnel differ from the normal sounds used for training.
Future work will investigate continual adaptation using newly collected audio clips verified as normal to update the normal-sound dictionary and LoRA parameters and recalibrate the detection threshold.

\section{Conclusion}
\label{sec:conclusion}
This paper proposed \method{}, which combines optimiza-tion-based separation with deep neural refinement to achieve accurate anomaly detection with listenable explanations.
The separated anomalous sound serves as the listenable explanation, and its energy determines the anomaly score.
Experiments on synthetic mixtures and real mixtures showed improved separation quality over baseline separators and detection performance competitive with the end-to-end baseline across the evaluated environments.
The results demonstrate that \method{} achieved the objective of accurate anomaly detection with listenable explanations in the evaluated environments.

\bibliographystyle{IEEEtranStyle_LimitNumAuthors}
\bibliography{1-reference}

\end{document}